\documentclass{article}
\usepackage{spconf,amsmath,graphicx,booktabs,threeparttable}
\usepackage{mathrsfs,bm}
\usepackage{subfigure}
\usepackage{url}

\title{Contrastive Regularization for Multimodal Emotion Recognition Using Audio and Text}
\name{Fan Qian, Jiqing Han}
\address{School of Computer Science and Technology, Harbin Institute of Technology, Harbin, China}

\begin{document}
%\ninept
%
\maketitle
\begin{abstract}
Speech emotion recognition is a challenge and an important step towards more natural human-computer interaction (HCI). The popular approach is multimodal emotion recognition based on model-level fusion, which means that the multimodal signals can be encoded to acquire embeddings, and then the embeddings are concatenated together for the final classification. However, due to the influence of noise or other factors, each modality does not always tend to the same emotional category, which affects the generalization of a model. In this paper, we propose a novel regularization method via contrastive learning for multimodal emotion recognition using audio and text. By introducing a discriminator to distinguish the difference between the same and different emotional pairs, we explicitly restrict the latent code of each modality to contain the same emotional information, so as to reduce the noise interference and get more discriminative representation. Experiments are performed on the standard IEMOCAP dataset for 4-class emotion recognition. The results show a significant improvement of 1.44\% and 1.53\% in terms of weighted accuracy (WA) and unweighted accuracy (UA) compared to the baseline system.
\end{abstract}
\section{INTRODUCTION}
\label{sec:introduction}

In the human-centered era, natural human-computer interaction (HCI) becomes more and more important. An ideal HCI system is required to detect the users emotion for enriching their experience, so as to provide better feedback. Many studies \cite{Soleymani2012MultimodalER,Aviezer2012BodyCN,Pantic2005AffectiveMH} show that a prominent automatic emotion recognition system should be multimodal since it is closer to the human sensor system. 

The core challenge of multimodal emotion recognition is how to integrate the emotional knowledge from various modalities. With the popularity of deep learning, model-level fusion has become the current mainstream method of multimodal emotion recognition, which is implemented by concatenating embeddings obtained from different modalities into a combined representation as the input of a classification model for end-to-end training \cite{Nicolaou2011ContinuousPO,He2015MultimodalAD,Huang2015AnIO}. It can capture complex non-linear multimodal feature correlations and make better advantages of deep neural networks.

However, inevitably, there are several samples that contain noise, which affects the complementarity between the modalities. Specifically, the noise may cause the various modalities of a sample to tend to different emotion categories. For example, when a model is well-trained on the 4-class emotional recognition task using audio and text, the audio score of a noisy sample indicates that the speech is neutral, but its text score indicates that the speech is happy. Although the model prediction based on scores summation of two modalities may be correct, there is a conflict between them. This phenomenon shows that, due to noise or other factors, the complementarity between various modalities of a sample is lost, resulting in the failure of learning good representation, which greatly affects the generalization of a model.

To alleviate the above problem, we propose a novel contrastive regularization method. Our motivation is that the learned embeddings should be restricted to contain the same emotional information as much as possible, so as to reduce the influence of noise and improve the system robustness. Contrastive learning \cite{anand2020contrastive} is a paradigm that leverages abundant negative example pairs to constrain information contained in the learned representation, which satisfies our demand well. More specifically, a discriminator is introduced to compare positive and negative example pairs as the explicit constraint with the same emotion, where the positive example pair is the embeddings obtained by encoding the audio and text from the same sample, and the negative example pair is two modal embeddings that one is the audio embedding in the positive pair, and the other is the text embedding from different samples as well as different emotions. Therefore, the complementarity between the two modalities is greatly enhanced.

\section{RELATED WORK}
\label{sec:relatedwork}
There are currently three main fusion methods in multimodal emotion recognition: \emph{feature-level} fusion, \emph{decision-level} fusion and \emph{model-level} fusion. Feature-level fusion \cite{Busso2004AnalysisOE,Zhao2018MultimodalMD,Han2019ImplicitFB}, a.k.a. early fusion, is implemented by simply concatenating features from multiple modalities into a combined vector as the input of a predictive model. However, this type of fusion method might suffer from the curse of dimensionality on small datasets, meanwhile larger dimensional features might stress computational resources during model training. Decision-level fusion \cite{Han2019ImplicitFB,Valstar2016AVEC2D,Han2017FromHT}, a.k.a. late fusion, on the contrary, combines predictions obtained from different modalities for a final prediction via a voting strategy. Nevertheless, it does not take into account the interaction between various modalities. As a compromise of early fusion and late fusion, model-level fusion which fuses the intermediate representations of different modalities may be the best choice for the fusion strategy \cite{Nicolaou2011ContinuousPO}.

However, most previous works based on model-level fusion for emotion recognition seldom take into account the effect of noise on modal complementarity. In fact, there are always samples whose multiple modalities tend to be different emotional categories due to the effect of noise, no matter how well the model is trained. Our intuition is that, by adding the regularization constraint of the same emotion explicitly, the performance of model-level fusion method can be improved. To this end, we draw inspirations from the idea of contrastive learning \cite{anand2020contrastive}, where the difference between positive and negative examples is captured, and propose a novel regularization method. In contrast to random negative sampling commonly used in self-supervised learning \cite{Becker1992SelforganizingNN}, we draw negative examples from different emotions to restrict the learned representation to contain the same emotional information and experimentally demonstrate the effectiveness of our proposed method.
\section{METHODOLOGY}
\label{sec:methodology}
\subsection{Model-level fusion}
\label{ssec:baseline}
We start by introducing the model-level fusion for emotion recognition using audio and text. The audio features $\bm{x}_a$ and text features $\bm{x}_t$ pass through an individual encoder to obtain the embedding vectors $\bm{e}_a$ and $\bm{e}_t$, respectively. Then the embedding vectors are concatenated together to classify emotions. We argue that, when embeddings are learned well enough, only a linear classifier is required, 
\begin{equation}
	\label{equ:concat}
	\bm{s}=\bm{W}\cdot[\bm{e}_a;\bm{e}_t]
\end{equation}
where $\bm{s}$ is the score vector and $\bm{W}$ is the weight matrix. 
Eq. (\ref{equ:concat}) can also be written as:
%\begin{equation}
%	\label{equ:add}
%	\begin{aligned}
%		\bm{s}_a=\bm{e}_{a}\bm{W}_1\\
%		\bm{s}_t=\bm{e}_{t}\bm{W}_2\\
%		\bm{s}=\bm{s}_a+\bm{s}_t
%	\end{aligned}
%\end{equation}
\begin{equation}
	\label{equ:add1}
	\bm{s}_a=\bm{W}_1\cdot \bm{e}_{a}
\end{equation}
\begin{equation}
	\label{equ:add2}
	\bm{s}_t=\bm{W}_2\cdot \bm{e}_{t}
\end{equation}
\begin{equation}
	\label{equ:add3}
	\bm{s}=\bm{s}_a+\bm{s}_t
\end{equation}
The weight matrix $\bm{W}$ is the concatenation of $\bm{W}_1$ and $\bm{W}_2$. In other words, the score of audio and text connected vector is the summation of their respective scores. For $i$-th sample, the classification loss is cross-entropy as follows,
\begin{equation}
	\label{equ:softmax}
	\mathcal{L}_1=-log\frac{exp(s_{ij})}{\sum_{c=1}^{C}exp(s_{ic})}
\end{equation}
where $C$ is the number of categories and $s_{ij}$ is the score of $i$-th sample $j$-th class.
\subsection{Contrastive regularization}
\label{ssec:method}
\begin{figure*}[tp] %htbp 代表图片插入位置的设置
	% \centering %图片居中
	\includegraphics[width=\linewidth]{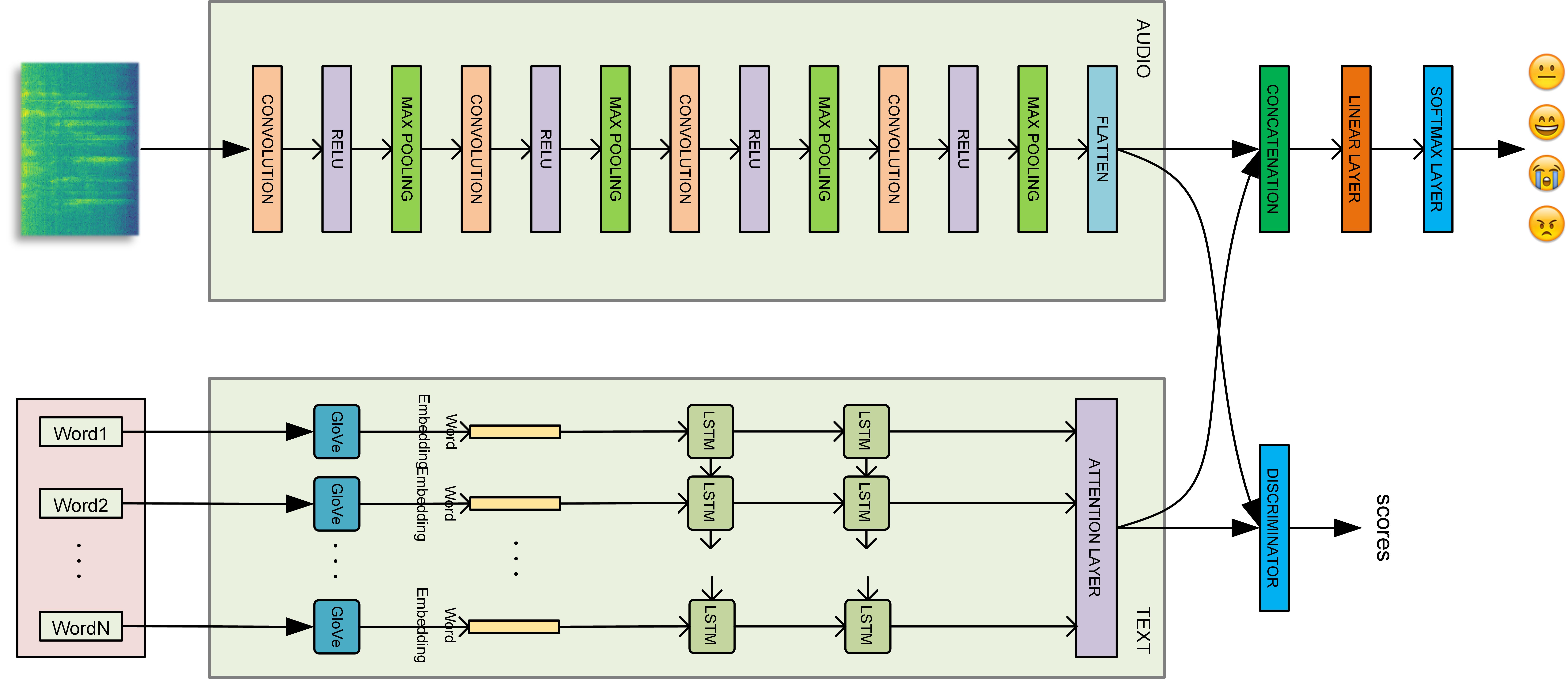} %[]中可选参数，可以设置图片的宽高
	%添加图体
	\caption{\centering Overview of our model framework}
	\label{fig:illustration}	
\end{figure*}
Due to the influence of noise or other factors, the categories corresponding to the largest scores of $\bm{s}_a$ and $\bm{s}_t$ are inconsistent. As a result, the complementarity between the two modalities is destroyed, which greatly affect the generalization of a model.

To overcome the above problem, we propose a contrastive regularization method and its model framework is shown in Fig. \ref{fig:illustration}. The upper block diagram represents a Convolution Neural Network (CNN) with the input of spectrograms as well as the output of audio embeddings. And the lower block diagram represents a Long-Short Time Memory (LSTM) with the input of word sequences as well as the output of text embeddings. The two output on the right side indicate the emotional classification loss and contrastive loss used, respectively. 

More specifically, a four-layers CNN is leveraged as the audio encoder for audio spectrogram features to capture the emotional information. Meanwhile, a two-layers LSTM with attention mechanism is used as the text encoder. The input of LSTM is GloVe \cite{Pennington2014GloveGV} word embeddings which embed a sequence of individual words from video segment transcripts into a sequence of word vectors that represent spoken text. The ensemble of CNN and LSTM is a backbone framework of both the baseline and our method. In addition, a discriminator is introduced and fed by either a positive example pair or negative example pairs and trained to separate them. Specifically, for each sample, we select its corresponding audio and text features as a positive pair, and at the same time select multiple text features with other different emotions as negative pairs. Through the comparison of positive and negative pairs, the learned representations are constrained to contain the same emotional information. Therefore, the noise is reduced and the loss of complementarity can be well mitigated.

For $i$-th sample, the contrastive loss is similar to the form of cross-entropy, which is called \emph{infoNCE} by \cite{Oord2018RepresentationLW},  
\begin{equation}
	\label{equ:contrastive}
	\mathcal{L}_2=-log\frac{exp(d(\bm{e}_{a}^{i},\bm{e}_{t}^{i}))}{exp(d(\bm{e}_{a}^{i},\bm{e}_{t}^{i}))+\sum_{n=1}^{N}exp(d(\bm{e}_{a}^{i},\bm{e}_{t}^{n^{-}}))}
\end{equation}
where $d(,)$ is the discriminator with the input of either positive example pair $(\bm{e}_{a}^{i}, \bm{e}_{t}^{i})$ or negative example pair $(\bm{e}_{a}^{i}, \bm{e}_{t}^{n^{-}})$, and $N$ is the number of negative pairs. In this work, a single hidden layer fully connected neural network is selected as the discriminator where the output is a real value, and $N$ is equal to batch size. 

The final loss function is as follows, 
\begin{equation}
	\label{equ:final_loss}
	\begin{aligned}
		\mathcal{L}&=(1-\alpha)\mathcal{L}_1+\alpha\mathcal{L}_2\\
	\end{aligned}
\end{equation}
where $\alpha$ is a hyperparameter that controls the strength of the regularization term. In this work, we use a grid search with a step size of 0.1 to find the optimal $\alpha$.
\section{EXPERIMENTS AND RESULTS}
\label{sec:experiment}
\begin{figure*}[tp]
	\begin{minipage}[t]{0.5\linewidth}
		\centering
		\includegraphics[width=\linewidth,trim={0 0 0 1.2cm},clip]{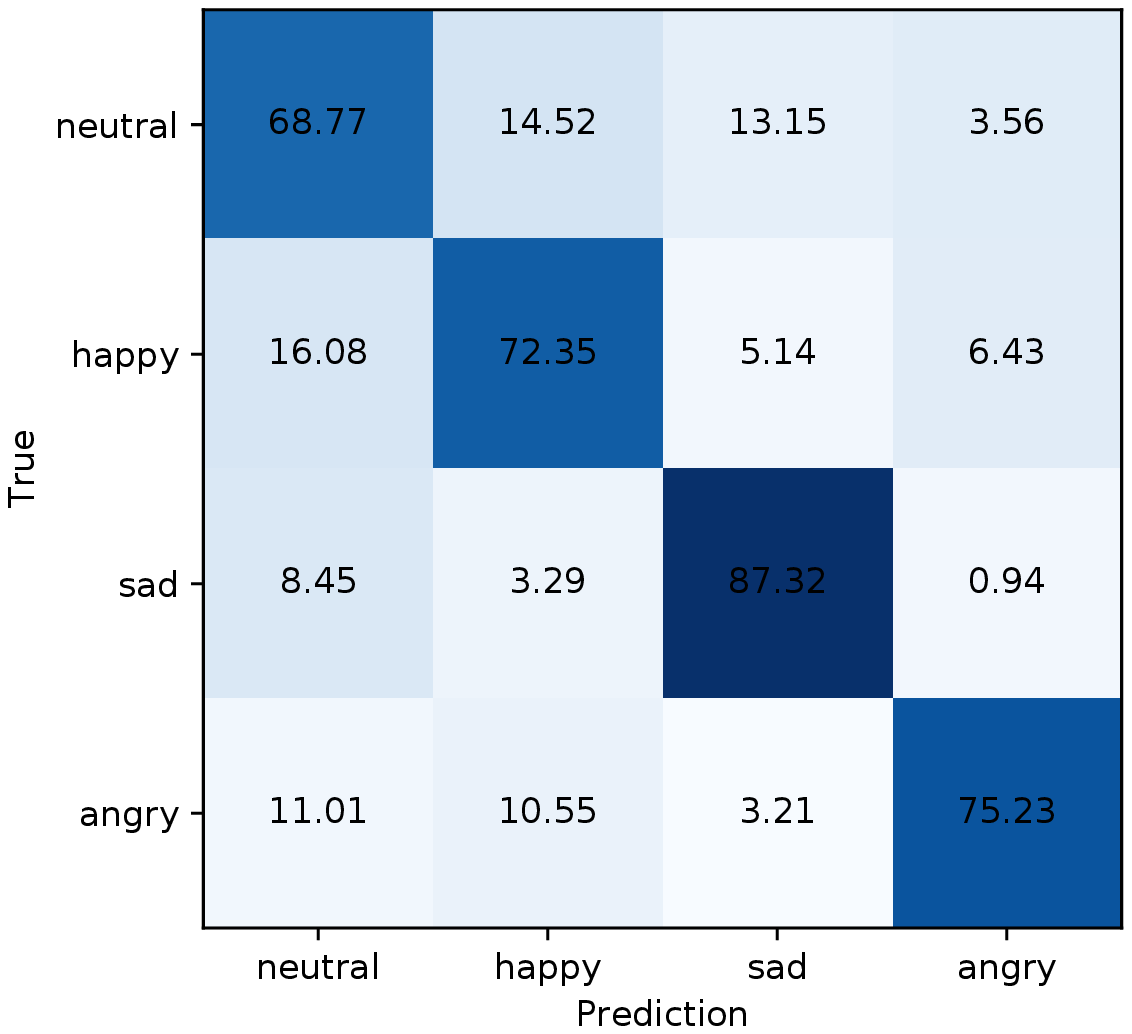}
	\end{minipage}
	\hfill
	\begin{minipage}[t]{0.5\linewidth}
		\centering
		\includegraphics[width=\linewidth,trim={0 0 0 1.2cm},clip]{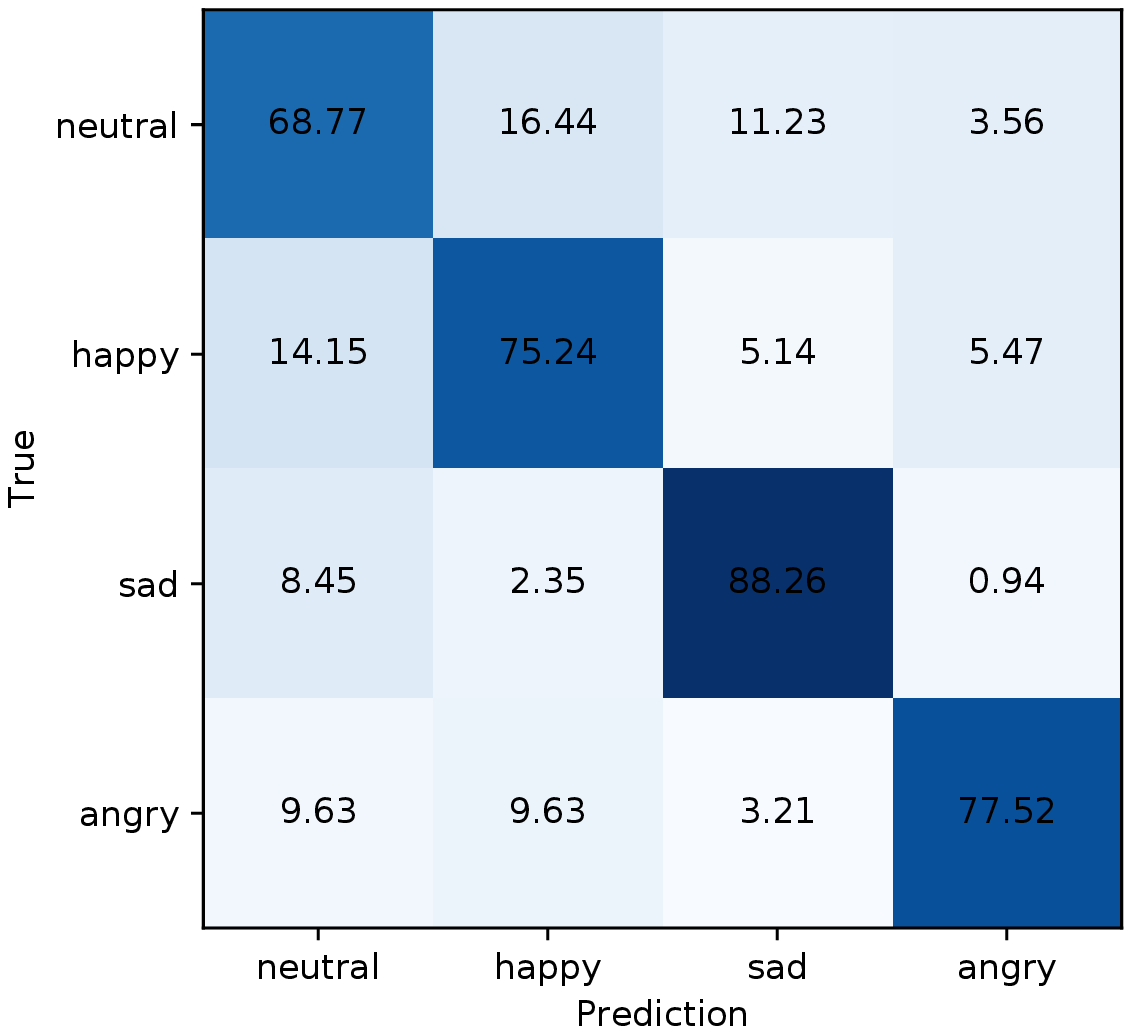}
	\end{minipage}
	\caption{\centering Confusion matrices of baseline and our method. Left and right figures represent baseline and our method respectively.}
	\label{fig:confusion_matrices}
\end{figure*}
\subsection{Database}
\label{ssec:database}
We use the IEMOCAP database \cite{Busso2008IEMOCAPIE} to evaluate the performance of different methods, which contains approximate 12 hours of audio-video recordings of dyadic interactions with 10 different speakers split in pair over 5 sessions in English. There are about 10000 utterances in the database that has been given emotion labels by at least 3 annotators. 12 unique raters annotate the database by choosing labels out of the 9 possible emotional labels per utterance. Emotional labels are happiness, anger, sadness, neutral, fear, surprise, frustration, excitement, and others.

In order to compare with majority of previous works \cite{Yoon2018MultimodalSE,Yoon2019SpeechER}, we select utterances annotated with four basic emotions: neutral, happiness, sadness and anger. Samples with excitement are merged with happiness resulting in a total of 5531 utterances (neutral: 1708, happiness: 1636, sadness: 1084, anger: 1103).
\subsection{Preprocessing}
\label{ssec:preprocessing}
For audio features, we extract spectrograms using the python toolkit, pylab \cite{Hunter2007MatplotlibA2}. We calculate a short-time Fourier transform with 256 FFT points and the hop length is 50\% of the number of FFT points. In addition, all audio utterances are divided into a uniform length of approximate one second. Thus, the input audio features are 128$\times$128 spectrograms.

For text features, we extract GloVe \cite{Pennington2014GloveGV} word embeddings utilizing gensim toolkit \cite{Rehurek2011GensimS}. The \emph{word2vec-google-news-300} \cite{Mikolov2013LinguisticRI}, a mass of pre-trained vectors trained on a part of the Google News dataset (about 100 billion words), is loaded. We split or pad each sentence into 30 words and acquire a 300-dimensional word embedding sequence of length 30.
\subsection{Experimental setup}
\label{ssec:setup}
For the CNN block used to process the spectrograms, a 3×3 convolution kernel is utilized with padding 1, and the max-pooling kernel is 2×2 with a 2×2 stride. Moreover, the activation function is ReLU and the output channels are all 8. The LSTM which is used to process word embedding sequences has 200 hidden nodes. It is a single-directional framework with two layers. An Adam optimizer is used with the learning rate of 0.001, and batch size is set to 64. The hyperparameter $\alpha$ is selected as 0.1 through a grid search in the range of [0.0, 1.0] with a step size of 0.1.

Following \cite{Yoon2018MultimodalSE,Yoon2019SpeechER}, we utilize 10-fold cross validation with a 8:1:1 split for training, validation and testing, respectively. The performance of the system is measured in terms of weighted accuracy (WA) and unweighted accuracy (UA) \cite{Han2014SpeechER}.
\subsection{Results and Analysis}
\label{ssec:results}

Table \ref{tab:result} presents the performance of our approach for emotion recognition compared with other methods. The baseline model is based on model-level fusion that concatenate embeddings obtained from audio and text modalities into a combined representation as the input of a linear classification model. DS-LSTM \cite{Wang2020SpeechER} is a dual-level model that predicts emotions based on both MFCC features and mel-spectrograms, and it is the state-of-the-art unimodal emotion recognition only using audio. MDRE \cite{Yoon2018MultimodalSE} encodes the information from audio and text sequences using dual recurrent neural network (RNN) and then combines the information from these sources to predict the emotion class. MHA-2 \cite{Yoon2019SpeechER} uses a multi-hop attention mechanism to exploit both textual and acoustic information in tandem.
\begin{table}[]
	\centering
	\caption{\centering Methods performance comparison on IEMOCAP database (\%)}	
	\label{tab:result}  	
	\begin{tabular}{ccc}  		
		\toprule[1pt]   		
		\textbf{Models} &\textbf{WA} &\textbf{UA} \\  		
		\midrule[1pt]
		Model-level fusion (Baseline) &74.62 &75.92 \\ 
		DS-LSTM \cite{Wang2020SpeechER} &72.7 &73.3 \\
		MDRE \cite{Yoon2018MultimodalSE} &71.8 &— \\
		MHA-2 \cite{Yoon2019SpeechER} &76.5 &77.6 \\ 		  		
		Contrastive regularization (Proposed) &76.06 &77.45 \\     				
		\bottomrule[1pt]  		
	\end{tabular}	
\end{table}

As illustrated in Table \ref{tab:result}, our proposed method outperforms DS-LSTM by 3.36\% and 4.15\% in terms of WA and UA, which further demonstrates the previous conclusion that the multimodal method is better than the unimodal method. In addition, it is worth noting that MHA-2, the state-of-the-art multimodal emotion recognition using audio and text, achieves 76.5\% in WA and 77.6\% in UA. Although our results are not directly comparable due to distinct model framework used and domain specialized optimizations, the idea behind our contrastive regularization method, which leverages abundant negative example pairs to acquire stronger representations, can be characterized as an additional contrastive loss and  serve as a regularization term. Therefore, our proposed method is more applicable than MHA-2 due to the limitation of multi-hop attention mechanism which is only used for audio and text sequential data.

More notably, the 1.44\% and 1.53\% improvement in terms of WA and UA compared to the baseline are realized by our proposed method. Meanwhile, as confusion matrices shown in Fig. \ref{fig:confusion_matrices}, our proposed method acquires the same accuracy 68.77\% compared to the baseline method in classifying neutral emotion and outperforms the baseline method in other three types of emotions. Moreover, the baseline model incorrectly classifies more examples of one emotion as another emotion than our proposed method, e.g., 16.08\% to 14.15\% for happy, 11.01\% to 9.63\% for angry, except for misclassification of neutral as happy (14.52\% to 16.44\%). For this type of misclassification, we guess that randomly sampling other different emotions against neutral class may sample more happy emotions due to the slight imbalance of the samples.

\section{CONCLUSIONS}
\label{sec:conclusions}
In this paper, we proposed a novel regularization method via contrastive learning for multimodal emotion recognition task. The proposed method is designed to constrain latent representations of audio and text to contain the same emotional information, so that it enhances the complementarity between the two modalities and improves the generalization of a model. Experiments show that the proposed method outperforms the baseline system in classifying the four emotion categories on IEMOCAP database by 1.44\% and 1.53\% in terms of WA and UA respectively and is competitive with the state-of-the-art system. 
\section{ACKNOWLEDGEMENTS}
\label{sec:knowledgements}
This research is supported by the National Natural Science Foundation of China under grant No.U1736210.
\vfill\pagebreak

% References should be produced using the bibtex program from suitable
% BiBTeX files (here: strings, refs, manuals). The IEEEbib.bst bibliography
% style file from IEEE produces unsorted bibliography list.
% -------------------------------------------------------------------------

\bibliographystyle{IEEEbib}
\bibliography{strings}

\end{document}